\documentclass[twocolumn,amsmath,amssymb]{revtex4-2}
\usepackage{graphicx}
\usepackage{dcolumn}
\usepackage{bm}

\input{macros.sty}
\usepackage{amsmath}
\usepackage{physics}

\newcommand{\ii}{{\rm i}}
\newcommand{\up}{{\uparrow}}
\newcommand{\down}{{\downarrow}}
\usepackage{xcolor}
\begin{document}

\preprint{APS/123-QED}

\title{Paramagnetic phases of strongly correlated ultracold fermions coupled to an optical cavity}

\author{Renan da Silva Souza}%
\email{souza@itp.uni-frankfurt.de}
\author{Youjiang Xu}
\author{Walter Hofstetter}
\email{hofstett@physik.uni-frankfurt.de}
\affiliation{%
Goethe-Universit\"at, Institut f\"ur Theoretische Physik, 60438 Frankfurt am
Main, Germany
}%

\date{\today}

\begin{abstract}
We numerically study a gas of two-component fermions coupled to a transversely pumped optical cavity and confined to a two-dimensional static square optical lattice.~In the dispersive regime, the steady state of the system is described by an extended Hubbard Hamiltonian with cavity-mediated long-range interactions.~Using real-space dynamical mean-field theory (RDMFT), we investigate the formation of the (superradiant) checkerboard density-wave phase.~Our analysis focuses on paramagnetic phases both at quarter and half filling.~At quarter filling, we find a reentrant homogeneous Fermi liquid to density wave phase transition with increasing temperature, which is due to the higher entropy of the ordered phase.~At half filling, in addition to the Fermi liquid to Mott insulator phase transition, marked by a vanishing quasiparticle residue at the Fermi level, we identify the transition into a density-wave phase.~Due to perfect Fermi surface nesting at half filling, we find that arbitrarily small long-range interactions destabilize the system towards the density-wave phase in the absence of short-range interactions.~By varying short- and long-range interactions at a fixed low temperature, we obtain the full phase diagram and identify a region of coexistence between the homogeneous Fermi liquid and Mott insulating phase with the density-wave phase.~In this region, we determine the thermodynamic phase transition by comparing the energies of the different RDMFT solutions.
\end{abstract}

\maketitle


\section{\label{sec:intro}Introduction}

Understanding the effect of nonlocal interactions in strongly correlated quantum many-body systems is a long-standing question in physics.~This question arises, for instance, when considering the long-range interactions of dipolar \cite{griesmaier2005bose, chicireanu2006simultaneous, lahaye2009physics, baranov2012condensed} and Rydberg-excited quantum gases \cite{labuhn2016tunable, zeiher2016many, bernien2017probing} or the Coulomb interaction \cite{hirsch1984charge,
lin1986condensation,
micnas1988superconductivity,
micnas1990superconductivity,
dagotto1994superconductivity,
chattopadhyay1997phase,
vojta1999charge,
pietig1999reentrant,
vojta2001phase,
hoang2002coherent,
calandra2002metal,
rosciszewski2003charge,
tong2004charge,
aichhorn2004charge,
merino2005quantum,
davoudi2006nearest,
merino2007nonlocal,
camjayi2008coulomb,
fratini2009unconventional,
amaricci2010extended,
merino2013emergent,
ayral2013screening,
huang2014extended,
hafermann2014collective,
van2014beyond,
terletska2017charge,
kapcia2017doping,
terletska2018charge} between electrons in a metal.~A unique approach to address this problem is to investigate a quantum gas coupled to a single-mode optical cavity driven by a pump laser \cite{mivehvar2021cavity}.~In this setting, the two-photon processes between the cavity and the pump mediate an effective global long-range interaction among the atoms.~This experimental setup has enabled the observation of the superradiant self-organization quantum phase transition, where a checkerboard density-wave pattern emerges in the atomic density simultaneously with steady-state superradiance marked by the macroscopic number of photons occupying the cavity mode which is caused by the coherent scattering of pump photons into the cavity.~This phenomenon was first observed in a transversely driven thermal gas coupled to an optical cavity \cite{black2003observation}, and later in a transversely driven atomic Bose–Einstein condensate (BEC) coupled to an optical cavity \cite{baumann2010dicke}.~Since then, significant experimental progress has been made with bosons coupled to optical cavities, including the investigation of symmetry breaking across the transition \cite{baumann2011exploring}, the measurement of the softening of the lowest-energy polariton mode \cite{mottl2012roton}, the observation of quench dynamics across the non-equilibrium phase transition \cite{klinder2015dynamical}, and the determination of the atomic dynamic structure factor from the power spectral density of the cavity radiation field \cite{landig2015measuring}.   

For bosons coupled to an optical cavity, the competition between onsite and cavity-mediated long-range interactions has been studied within the extended Bose–Hubbard model \cite{PhysRevA.87.051604,dogra2016phase,niederle2016ultracold,panas2017spectral, liao2018theoretical, carl2023phases}, which has also been realized experimentally \cite{klinder2015observation,landig2016quantum,hruby2018metastability}.~In particular, for single-component bosons in the strong-interaction regime, a region of metastability where Mott-insulating and density-wave phases coexist was predicted \cite{landig2016quantum,panas2017spectral} and later confirmed experimentally \cite{hruby2018metastability}.

In recent experiments the effect of cavity-mediated long-range interactions on two-component spinful ultracold Fermi gases of  $^6$Li atoms was studied \cite{roux2020strongly,roux2021cavity,konishi2021universal,zhang2021observation,helson2023density,zwettler2025nonequilibrium,Buhler_2025_Microscopy}.~The self-organization of the atomic density and the associated superradiance was first realized for a noninteracting ultracold Fermi gas \cite{zhang2021observation} and more recently for a unitary Fermi gas with attractive contact interactions due to $s$-wave scattering, where the scattering length was tuned across the crossover from a BEC of tightly bound molecules to a Bardeen-Cooper-Schrieffer (BCS) superfluid of weakly bound Cooper pairs \cite{helson2023density}.~One of the main novel features of the fermionic superradiance compared to its bosonic counterpart is its strong dependency on the atomic density.~As demonstrated theoretically for spinless fermions without short-range interactions \cite{keeling2014fermionic,piazza2014umklapp,chen2014superradiance}, the Fermi surface nesting effect, where portions of the Fermi surface are connected by a single wave vector matching the cavity photon momentum, can significantly decrease the threshold for atomic self-organization at specific fillings.~By contrast, for other fillings, the Pauli blocking effect suppresses this self-organization, since the momentum states coupled by photon absorption are already occupied.

One possibility of investigating the Fermi surface effect in greater detail is to confine the system to a two-dimensional static square optical lattice.~For sufficiently deep lattices and at low fillings, the atomic system can be described by a Fermi-Hubbard model \cite{hofstetter2018quantum}.~In this setting, the Fermi surface of the two component Fermi gas exhibits perfect nesting \cite{hafermann2014collective}.~Such a lattice further allows for tunable onsite Hubbard interactions between the two fermionic components.

In the fermionic case, extended Hubbard models have been widely used to describe strongly correlated electrons in solids, where the Coulomb interaction is only partially screened.~Considering nearest-neighbor interactions, this model has been investigated using a variety of techniques, including among others, Hartree-Fock mean-field \cite{micnas1988superconductivity,micnas1990superconductivity,chattopadhyay1997phase, rosciszewski2003charge}, coherent potential approximation \cite{hoang2002coherent}, variational cluster approach \cite{aichhorn2004charge}, exact diagonalization \cite{calandra2002metal, merino2005quantum, fratini2009unconventional}, the two-particle self-consistent approach \cite{davoudi2006nearest}, Monte Carlo simulations \cite{hirsch1984charge,lin1986condensation,dagotto1994superconductivity}, density-matrix renormalization group \cite{vojta1999charge,vojta2001phase}, and dynamical mean-field theory (DMFT) \cite{pietig1999reentrant,tong2004charge,camjayi2008coulomb,amaricci2010extended,merino2013emergent,kapcia2017doping} and its extensions \cite{merino2007nonlocal,ayral2013screening,ayral2013screening,huang2014extended,hafermann2014collective,van2014beyond,terletska2017charge,terletska2018charge}.~At quarter filling these models are particularly relevant, as they capture the interplay between charge order and superconductivity observed in layered molecular crystals such as bis-(ethylenedithia-tetrathiafulvalene) (BEDT-TTF) \cite{merino2001superconductivity,mckenzie2001charge,merino2007nonlocal}.~Using standard DMFT on a Bethe lattice, Ref.~\cite{pietig1999reentrant} demonstrated that nearest-neighbor repulsion leads to reentrant behavior in the transition from a homogeneous metallic phase to a charge-ordered phase upon increasing temperature.~More recently, Ref.~\cite{kapcia2017doping} studied the same model at general filling and zero temperature, identifying Fermi-liquid, Mott-insulating, and charge-ordered phases, as well as extended metastable regions at half filling where charge-ordered and homogeneous solutions coexist.~The realization of cavity-mediated long-range interactions in ultracold Fermi gases coupled to an optical cavity provides the possibility of observing similar physics in a highly controllable setting.

For a Fermi-Hubbard chain coupled to a transversely pumped cavity, exact diagonalization revealed that tuning the pump polarization allows stabilization of diverse phases, including density waves, antiferromagnetic insulators, pair superfluids, and pair density waves \cite{camacho2017quantum}.~Recently, Refs.~\cite{tolle2025fluctuation,tolle2025steady} analyzed the same system including fluctuations beyond the mean-field atom–cavity coupling, revealing a regime of bistability in the superradiant, self-organized phase.~Despite these advances, the interplay between cavity-mediated long-range interactions and onsite Hubbard repulsion in fermionic lattice systems remains largely unexplored.~In particular, it is an open question whether the transition between the Mott insulator and the density-wave insulator is of first order, as has been observed for bosons coupled to a cavity \cite{panas2017spectral,hruby2018metastability}.

\begin{figure}[!]
\includegraphics[width=1\linewidth]{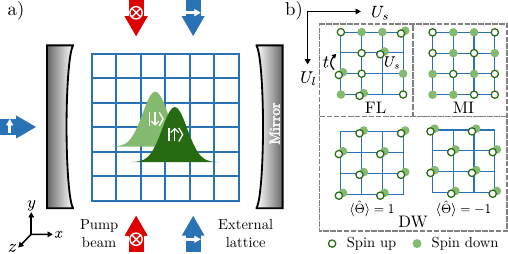}
\caption{\label{fig:sys}a) Schematic representation of a two-component Fermi gas (with the two components (hyperfine states) denoted as $\ket{\up}$, $\ket{\down}$) confined to a static square optical lattice inside a single-mode optical cavity.~The atoms are transversely driven by a coherent pump-laser field.~b) Schematic illustration of the atomic configuration of the different quantum phases we observe at half filling.~For weak long-range interactions $U_l$, we observe a phase transition from a Fermi-liquid phase (FL) to a Mott-insulating phase (MI) as the short-range interaction $U_s$ is increased.~For fixed $U_s$, increasing $U_l$ drives the system into a density-wave insulating phase (DW), in which the atoms self-organize, occupying either the even or the odd sublattice.~In this phase the expectation value of the occupation imbalance operator $\langle\hat{\Theta}\rangle$ [Eq.\,\eqref{eq7}] asymptotically approaches $+1$ for a checkerboard pattern with occupied even sites and $-1$ for occupied odd sites.}
\end{figure}

Motivated by these considerations, we consider a two‐component Fermi gas in a two-dimensional static square optical lattice dispersively coupled to a single-mode optical cavity and transversely pumped by a coherent laser field [Fig.\,\ref{fig:sys}(a)].~The system is described by an extended Hubbard model with cavity‐mediated density-density long‐range interactions.~We consider a balanced configuration, where the densities of the two fermionic components are equal and no magnetic order emerges in the system.~The global range of the cavity-mediated interactions justifies their static mean-field decoupling, as demonstrated for bosonic systems coupled to a cavity in \cite{piazza2013bose}.~This approach has proven to be effective in characterizing the phase transitions in such systems  \cite{PhysRevA.87.051604,dogra2016phase,niederle2016ultracold,panas2017spectral}.~We analyze the resulting problem using real-space dynamical mean-field theory (RDMFT) \cite{snoek2008antiferromagnetic}, a method that allows us to characterize the competing phases and their transitions across parameter regimes.~We study the phase transitions both at quarter- and half-filling characterizing the competition between self-organized checkerboard density-wave phase, homogeneous paramagnetic metallic (Fermi liquid) and Mott insulating phases [Fig.\,\ref{fig:sys}(b)].

In the following, we derive the effective Hamiltonian in Sec. \ref{sec:model}, then describe the RDMFT method in Sec. \ref{sec:methods}, and present and discuss our results in Sec. \ref{sec:results}.~Final remarks are presented in Sec. \ref{sec:conc}.

\section{\label{sec:model}Fermi-Hubbard model with long-range interactions}

We consider a balanced two-component ultracold Fermi gas with contact interactions arising from $s$-wave scattering, coupled to a single-mode high-finesse optical cavity.~The quantized cavity mode of frequency $\omega_c$ and wave number $k_c=k$ is $\lambda$-periodic along the $x$ direction.~The Fermi gas is driven by a coherent pump laser of frequency $\omega_p$ and wave number $k_p\approx k$ along the $y$ axis.~The system is confined to a two-dimensional layer of a cubic optical lattice, which is deep along the $z$ direction, and experiences a static square optical lattice potential with wave vector $k_{\rm latt}=k$ and $N_s$ sites in the $xy$ plane [see Fig.\,\ref{fig:sys}(a)], similar to the experimental setting realized in \cite{landig2016quantum}.~The two components, denoted as $\ket{\up}$ and $\ket{\down}$, correspond to the two lowest hyperfine states of $^{6}$Li, as experimentally realized in \cite{roux2020strongly,zhang2021observation}, where in the Paschen-Back regime, the transition $\ket*{2S_{1/2},m_J=-1/2}\rightarrow\ket*{2P_{3/2},m_J=-3/2}$ serves as a two-level system with resonant frequency $\omega_a$ for each component.

In the frame rotating with frequency $\omega_p$, the effective Hamiltonian describing the system is $\hat{H}_{\rm A}-\Delta_c\hat{a}^\dagger\hat{a}$, where $\hat{a}$ is the bosonic cavity mode annihilation operator and $\Delta_c=\omega_p-\omega_c$ is the pump-cavity detuning.~We separate the Hamiltonian describing the fermionic atoms into a noninteracting part $\hat{H}_0$ and a two-body interaction term $\hat{H}_{\rm int}$, $\hat{H}_{\rm A}=\hat{H}_0+\hat{H}_{\rm int}$.~If the atom-pump detuning $\Delta_a=\omega_p-\omega_a$ is large, the dynamics of the atomic excited state can be adiabatically eliminated \cite{maschler2005cold,maschler2008ultracold}.~In this regime, the free-fermion Hamiltonian reads
\begin{equation}
    \hat{H}_0=\sum_\sigma\int\,\dd^3 x\hat{\Psi}^\dagger_\sigma(\bm{x})\hat{h}_0\hat{\Psi}^{\vphantom\dagger}_\sigma(\bm{x}),
\end{equation}
\begin{equation}
    \hat{h}_0=-\frac{\hbar^2}{2m}\grad^2+\hat{V}_{\rm eff}(\bm{x})-\mu_\sigma,
\end{equation}
where $\hat{\Psi}^{\vphantom\dagger}_{\sigma^{\vphantom\prime}}(\bm{x})$ is the fermionic annihilation field operator of an atom in state $\sigma\in\{\up, \down\}$.~The two fermionic components experience an effective potential
\begin{equation}\label{eq3}
    \hat{V}_{\rm eff}(\bm{x})=V_{\rm latt}(\bm{x})+\hbar[V(\bm{x})+U(\bm{x})\hat{a}^\dagger\hat{a}+\eta(\bm{x})(\hat{a}+\hat{a}^\dagger)],
\end{equation}
where $V_{\rm latt}(\bm{x})=V_{2\rm D}[\cos^2(kx)+\cos^2(ky)]$ is the external static square optical lattice with depth $V_{2\rm D}$, $V(\bm{x})=V_0\cos^2(ky)$ is the pump lattice with depth $\hbar V_0=\hbar\Omega_0^2/\Delta_a$, where $\Omega_0$ is the maximum Rabi frequency of the pump laser, $U(\bm{x})=U_0\cos^2(kx)$ is the cavity potential with depth per photon $\hbar U_0=\hbar\mathcal{G}_0^2/\Delta_a$, where $\mathcal{G}_0$ is the maximum coupling strength of the atom to the cavity, and $\eta(\bm{x})=\eta_0\cos(kx)\cos(ky)$ is the checkerboard lattice generated by cavity-pump coupling with $\hbar\eta_0=\hbar\Omega_0\mathcal{G}_0/\Delta_a$ being the maximum two-photon Rabi frequency.~Furthermore, the two-body interatomic interaction is described by the Hamiltonian
\begin{equation}
    \hat{H}_{\rm int}=g\int\,\dd^3 x\hat{\Psi}^\dagger_{\up}(\bm{x})\hat{\Psi}^\dagger_{\down}(\bm{x})\hat{\Psi}^{\vphantom\dagger}_{\down}(\bm{x})\hat{\Psi}^{\vphantom\dagger}_{\up}(\bm{x}),
\end{equation}
where $g=4\pi a_s\hbar^2/m$ with $a_s$ being the $s$-wave scattering length.

When the static optical lattice $V_{\rm latt}$ is sufficiently deep, the atomic field operators can be expanded in terms of the lowest band Wannier functions as $\hat{\Psi}^{\vphantom\dagger}_{\sigma^{\vphantom\prime}}(\bm{x})=\sum_{\bm{i}}w_{\bm{i}}(\bm{x})\hat{c}_{\bm{i}\sigma}$, where $\hat{c}_{\bm{i}\sigma}$ is the annihilation operator for a fermion in spin state $\sigma$ at site $\bm{i}=(i_x,i_y)$ and $w_{\bm{i}}(\bm{x})$ is the corresponding Wannier function \cite{hofstetter2018quantum}.~Furthermore, we assume that the detuning $\Delta_c$ and the cavity damping rate $\kappa$ are much larger than the recoil frequency $\omega_r=\hbar k_c^2/2m$ of an atom of mass $m$, making the cavity field dynamics evolve on a much faster timescale than the center of mass motion of the atoms, thus following the latter adiabatically \cite{maschler2008ultracold}.~Consequently, the dynamics of the cavity mode can be simplified by using a steady-state solution and inserting it into \eqref{eq3}, which finally leads to the following extended Fermi-Hubbard Hamiltonian
\begin{equation}\label{eq5}
    \hat{H}=\hat{H}_{\rm HB}-\frac{U_l}{2}\Big(\sum_{\bm{i}\sigma}(-1)^{\abs{\bm{i}}}\hat{n}_{\bm{i}\sigma}\Big)^2,
\end{equation}
\begin{equation}
    \begin{aligned}
        \hat{H}_{\rm HB}=&-\sum_{\bm{ij}\sigma}t_{\bm{ij}}\hat{c}^\dagger_{\bm{i}\sigma}\hat{c}^{\vphantom\dagger}_{\bm{j}\sigma}+U_s\sum_{\bm{i}}\hat{n}_{\bm{i}\up}\hat{n}_{\bm{i}\down}
        -\sum_{\bm{i}\sigma}\tilde{\mu}_\sigma\hat{n}_{\bm{i}\sigma},
    \end{aligned}
\end{equation}
where we have defined $\hat{n}_{\bm{i}\sigma}=\hat{c}^\dagger_{\bm{i}\sigma}\hat{c}^{\vphantom\dagger}_{\bm{i}\sigma}$.~The first term of $\hat{H}_{\rm HB}$ represents the tunneling amplitude, with $t_{\bm{ij}}=t$ if $\bm{i,j}$ are nearest neighbors and zero otherwise, while the second term corresponds to the Hubbard onsite interaction of strength $U_s$.~The chemical potential $\tilde{\mu}_\sigma=\mu_{\sigma}-\hbar V_0 B$ is displaced with a shift proportional to the overlap of the pump lattice with the Wannier functions $B=\int\dd^3x\,|w_{\bm{i}}(\bm{x})|^2\cos^2(ky)$.~The second term in \eqref{eq5} corresponds to the cavity-mediated long-range interactions and is proportional to the square of the occupation imbalance operator 
\begin{equation}\label{eq7}
   \hat{\Theta}\equiv\frac{1}{N_s}\sum_{\bm{i}\sigma}(-1)^{\abs{\bm{i}}}\hat{n}_{\bm{i}\sigma},
\end{equation}
where $\abs{\bm{i}}=i_x+i_y$, multiplied by the density-density interaction strength $U_l=-4\hbar A^2\Delta_c\eta_0^2/(\Delta_c^2+\kappa^2)$, where $A=\int\dd^3x\,|w_{\bm{i}}(\bm{x})|^2\eta(\bm{x})/\eta_0$.~Notably, unlike standard extended Hubbard models with nearest-neighbor interactions, the density-density interactions in \eqref{eq5} are of infinite range.~By tuning the depth of the static optical lattice $V_{2\rm D}$ and the pump-cavity detuning $\Delta_c$, the strength of the tunneling $t$, short- and long-range interactions $U_s$ and $U_l$ can be adjusted \cite{jaksch1998cold,zwerger2003mott,bloch2008many,hoffmann2009visibility,giorgini2008theory,landig2016quantum}.~We assume a red relative detuning $\Delta_c<0$, such that $U_l>0$.~As demonstrated in \cite{piazza2014quantum}, the relaxation rate of the atomic ensemble is suppressed when the ratio $\omega_r/\kappa$ becomes small, which justifies using thermal-equilibrium approaches to describe the steady state of the atomic system.~In the following we describe the RDMFT method used in our analysis.\

\section{\label{sec:methods}Real-Space Dynamical Mean-Field Theory}

The infinite range of the cavity-mediated long-range interactions in \eqref{eq5} justifies applying a static mean-field decoupling $\hat{\Theta}^2\approx2\langle\hat{\Theta}\rangle\hat{\Theta}-\langle\hat{\Theta}\rangle^2$, where the expectation value $\langle\hat{\Theta}\rangle$ has to be determined self-consistently within an iterative algorithm.~This expectation value measures the imbalance between the occupations of even ($\abs{\bm{i}}=2p$ with $p\in\mathbb{N}_0$) and odd ($\abs{\bm{i}}=2p+1$ with $p\in\mathbb{N}_0$) lattice sites, thus serving as an order parameter for the emergence of the checkerboard density order.~This static mean-field decoupling of the cavity-mediated long-range interaction has been applied to ultracold bosons coupled to a cavity and turned out to be a good approximation to capture the quantum phase transitions present in such systems \cite{PhysRevA.87.051604,dogra2016phase,niederle2016ultracold,landig2016quantum,panas2017spectral}.~Within this approximation the density-density long-range interaction reduces to an effective local staggered potential, and the system can be described by the mean-field Hamiltonian  
\begin{equation}\label{efH}
    \hat{H}_{\rm MF}=\hat{H}_{\rm HB}-N_sU_l\langle\hat{\Theta}\rangle\sum_{\bm{i}\sigma}(-1)^{\abs{\bm{i}}}\hat{n}_{\bm{i}\sigma} + N_s^2\frac{U_l}{2}\langle\hat{\Theta}\rangle^2.
\end{equation}

To calculate the equilibrium phase diagram of the system described by $\hat{H}_{\rm MF}$, we use real-space dynamical mean-field theory (RDMFT), which provides a non-perturbative treatment of strongly correlated lattice systems \cite{snoek2008antiferromagnetic} by fully incorporating the dynamics of local quantum fluctuations at each site which are induced by short-range interactions.~The RDMFT framework is an extension of standard dynamical mean-field theory (DMFT) \cite{georges1996dynamical} where the strongly correlated $N_s$ site lattice problem is mapped to a set of $N_s$ single-impurity Anderson models, which are coupled by a self-consistency condition \cite{snoek2008antiferromagnetic,helmes2008mott}.~The self consistency in this approach is formulated in real space and relies on assuming a position-dependent local self-energy.~A derivation of (real space) DMFT can be obtained by first writing the partition function for the system in the coherent-state path-integral representation and choosing a single lattice site $\bm{i}$, which is denoted as the impurity site, and proceeding by integrating out all other degrees of freedom \cite{georges1996dynamical}.~In the limit of an infinite lattice coordination number $\mathsf{z}$, with the proper rescaling of the hopping amplitude as $t\rightarrow t^*/\sqrt{\mathsf{z}}$, the effective impurity problem of lattice site $\bm{i}$ can be described by the effective action
\begin{equation}\label{eq8}
    \begin{aligned}
    S_{\rm eff}^{(\bm{i})}=&-\int^\beta_0\dd\tau\int^\beta_0\dd\tau^\prime\sum_\sigma c^*_{\bm{i}\sigma}(\tau)\mathcal{G}_\sigma^{(\bm{i})}(\tau-\tau^\prime)^{-1}c^{\vphantom *}_{\bm{i}\sigma}(\tau)
    \\
    &+U_s\int^\beta_0\dd\tau n_{\bm{i}\up}(\tau)n_{\bm{i}\down}(\tau),
    \end{aligned}
\end{equation}\par\noindent
where $\tau\in[0,\beta]$ is imaginary time, $\beta=1/T$ is the inverse temperature, and we assume $\hbar=k_{\rm B}=1$.~The fields $c^{\vphantom *}_{\bm{i}\sigma}(\tau)$ and $c^*_{\bm{i}\sigma}(\tau)$ are Grassmann variables and we define $n^{\vphantom *}_{\bm{i}\sigma}(\tau)=c^*_{\bm{i}\sigma}(\tau)c^{\vphantom *}_{\bm{i}\sigma}(\tau)$.~The Weiss mean-field $\mathcal{G}_\sigma^{(\bm{i})}(\tau-\tau^\prime)$ represents the effect of all other sites on site $\bm{i}$.~To solve the impurity problem, the effective action \eqref{eq8} is mapped onto an effective Anderson impurity Hamiltonian, chosen to be of the form
\begin{equation}\label{eq:H_AIM}
    \begin{aligned}
    \hat{H}^{(\bm{i})}=&-\sum_\sigma \gamma_{\sigma}^{(\bm{i})}\hat{d}_{\sigma}^{\dagger}\hat{d}_{\sigma}^{\vphantom\dagger}+U_s\hat{d}_{\up}^{\dagger}\hat{d}_{\up}^{\vphantom\dagger}\hat{d}_{\down}^{\dagger}\hat{d}_{\down}^{\vphantom\dagger}
    \\
    &+\sum_{b\sigma}\epsilon_{b\sigma}^{(\bm{i})}\hat{c}_{b\sigma}^\dagger\hat{c}_{b\sigma}^{\vphantom\dagger}+\sum_{b\sigma}(v_{b\sigma}^{(\bm{i})}\hat{d}_{\sigma}^{\dagger}\hat{c}_{b\sigma}^{\vphantom\dagger}+\rm h.c.),
    \end{aligned}
\end{equation}\par\noindent
where the operators $\hat{d}_{\sigma}^{(\dagger)}$ act on the impurity site and the operators $\hat{c}_{b\sigma}^{(\dagger)}$ act on the bath orbitals labeled by the index $b$.~The impurity orbital energy is defined as $-\gamma_{\sigma}^{(\bm{i})}=-\tilde{\mu}_\sigma-(-1)^{\abs{\bm{i}}}N_sU_l\langle\hat{\Theta}\rangle$.~For a given Weiss mean-field $\mathcal{G}_\sigma^{(\bm{i})}$ and a given initial occupation imbalance (usually set to $|\langle\hat{\Theta}\rangle|\approx0.2$ in our calculations), the algorithm proceeds by fitting the energies of the bath orbitals $\epsilon_{b\sigma}^{(\bm{i})}$ and the couplings between bath and impurity orbitals $v_{b\sigma}^{(\bm{i})}$ to reproduce the first term of the effective local action \eqref{eq8} via
\begin{equation}\label{eq:hybr}
    \mathcal{G}_\sigma^{(\bm{i})}(\ii\omega_m)^{-1}=\ii\omega_m+\gamma_{\sigma}^{(\bm{i})}+\sum_b\frac{\lvert v_{b\sigma}^{(\bm{i})}\rvert^2}{\epsilon_{b\sigma}^{(\bm{i})}-\ii\omega_m},
\end{equation}
where $\omega_m=(2m+1)\pi/\beta$ are fermionic Matsubara frequencies with $m\in\mathbb{Z}$.~Using the resulting values of $\epsilon_{b\sigma}^{(\bm{i})}$ and $v_{b\sigma}^{(\bm{i})}$, we construct the effective impurity Hamiltonian \eqref{eq:H_AIM} and solve it by exact diagonalization, where we have to assume that the number of bath orbitals is finite to limit the Hilbert-space dimension (usually five).~This method has the advantage of giving access to the real frequency spectrum without the need for numerical analytic continuation.~By calculating the eigenstates $\ket*{n^{(\bm{i})}}$ and eigenenergies $E_n^{(\bm{i})}$ of \eqref{eq:H_AIM}, we obtain the impurity Green's function in the Lehmann representation
\begin{equation}
    G_{\sigma}^{(\bm{i})}(\ii\omega_m) =\frac{1}{\mathcal{Z}^{(\bm{i})}}\sum_{l,n} 
\frac{(e^{-\beta E_l^{(\bm{i})}} + e^{-\beta E_n^{(\bm{i})}})\lvert \langle l^{(\bm{i})} | d_{\sigma}^\dagger | n^{(\bm{i})} \rangle \rvert^2}{\ii\omega_m + E_l^{(\bm{i})} - E_n^{(\bm{i})}},
\end{equation}
and use the local Dyson equation to compute the position dependent local self-energy in real space 
\begin{equation}\label{eq:locDy}
\Sigma_{\bm{i},\sigma}(\ii\omega_m)=\mathcal{G}^{(\bm{i})}_\sigma(\ii\omega_m)^{-1}-G_{\sigma}^{(\bm{i})}(\ii\omega_m)^{-1}.  
\end{equation}
Here, $\mathcal{Z}^{(\bm{i})}=\sum_ne^{-\beta E_n^{(\bm{i})}}$ is the partition function of the impurity model.~The interacting lattice Green's function is then calculated from the lattice Dyson equation 
\begin{equation}\label{eq:lattDy}
[\vb{G}_\sigma(\ii\omega_m)]^{-1}=(\tilde{\mu}_\sigma+\ii\omega_m)\vb{1}+\vb{t}+\vb{V}-\vb{\Sigma}_\sigma(\ii\omega_m), 
\end{equation}
where we use bold symbols to represent real space matrices labeled by lattice site indices $\bm{i,j}$.~Here, $\vb{1}$ is the identity matrix, $\vb{t}$ is the hopping matrix with elements $t_{\bm{ij}}$, and $\vb{V}$ with matrix elements $V_{\bm{ij}}=\delta_{\bm{ij}}V_{\bm{i}}=\delta_{\bm{ij}}(-1)^{\abs{\bm{i}}}N_sU_l\langle\hat{\Theta}\rangle$ represents a spatially varying potential, which arises due to the mean-field decoupling of the cavity-mediated long-range interactions.~The local self-energy is diagonal in real space with matrix elements $[\vb{\Sigma}_{\sigma}(\ii\omega_m)]_{\bm{ij}}=\delta_{\bm{ij}}\Sigma_{\bm{i},\sigma}(\ii\omega_m)$.~Self consistency is achieved by equating the diagonal elements of the lattice Green's function with the local impurity Green's function, $G_{\sigma}^{(\bm{i})}(\ii\omega_m)=[\vb{G}_\sigma(\ii\omega_m)]_{\bm{ii}}$, and subsequently calculating the Weiss mean field from the local Dyson equation \eqref{eq:locDy}.~Solving the impurity problem allows us to calculate the local on-site occupations, which we use to obtain the occupation imbalance $\langle\hat{\Theta}\rangle$ updated in the simulation after each RDMFT iteration.~Methods based on DMFT are exact only in the limit of infinite dimensions (or equivalently, large coordination number).~However, they provide a reliable approximation for studying the phases of finite-dimensional lattice systems such as the one considered here, as has been consistently demonstrated in numerous applications (see, for instance, Refs.\,\cite{pietig1999reentrant,tong2004charge,camjayi2008coulomb,amaricci2010extended,merino2013emergent,kapcia2017doping}).

In principle, for a translationally invariant model, Eq.~\eqref{eq:lattDy} could be solved in the thermodynamic limit by Fourier transformation.~Within RDMFT, however, translational invariance is not assumed a priory and solutions exhibiting different types of order, such as staggered magnetization, phase separation, and density order, arise naturally without imposing additional symmetry constraints.~By contrast, it requires more computational resources than standard DMFT due to the inversion of $N_s\times N_s$ matrices in \eqref{eq:lattDy} and because for every lattice site an effective quantum impurity problem needs to be solved.~The latter task can be significantly simplified using insight from the expected type of symmetry breaking.~The transition to the checkerboard density-wave phase corresponds to a $\mathbb{Z}_2$ parity-symmetry breaking \cite{mivehvar2021cavity} towards a bipartite structure with finite occupation imbalance $\langle\hat{\Theta}\rangle$ between even and odd lattice sites.~After confirming this translational-symmetry breaking with unbiased RDMFT for a system of size $N_s=8\times 8$, we have applied a unit-cell procedure \cite{sotnikov2015critical} which assumes that all sites within the same sublattice (even or odd) are equivalent.~This reduces the number of different quantum impurity problems to one per sublattice, while the lattice Green's function is still computed for the full system via \eqref{eq:lattDy}.~This allowed us to simulate lattices of size $N_s=32\times 32$ with reduced computational cost, thus minimizing finite-size effects.

\section{\label{sec:results}Results and discussion}

As previously discussed, the occupation imbalance $\langle\hat{\Theta}\rangle$ serves as an order parameter for the checkerboard density-wave phase marking the self-organization of the atoms on the even or odd lattice sites.~On the other hand, the paramagnetic Mott transition can be characterized by the quasiparticle residue which measures the spectral weight at the Fermi level and becomes zero at the critical point marking the absence of quasiparticles in the Mott insulator phase and the vanishing of the metallic DMFT solution \cite{georges1996dynamical}.~For each sublattice, we obtain the quasiparticle residue from the converged self-energy via \cite{huang2014extended,kumar2016interaction}
\begin{equation}
    Z_{\vb{e},\vb{o}}=\qty(1-\frac{{\rm Im\Sigma_{\vb{e},\vb{o}}(\ii\omega_0)}}{\omega_0})^{-1},
\end{equation}
where the indices $\vb{e}$ and $\vb{o}$ correspond to sites in the even and odd sublattices, respectively, and we have suppressed the spin index assuming $\Sigma_{\bm{i},\up}(\ii\omega_m)=\Sigma_{\bm{i},\down}(\ii\omega_m)=\Sigma_{\bm{i}}(\ii\omega_m)$, which is valid for the paramagnetic phases studied here.~Overall we characterize the emerging phases of the systems as follows:
\begin{itemize}
    \item[1)] The homogeneous Fermi-liquid phase (FL) is characterized by a finite value of $Z_{\vb{e},\vb{o}}$ corresponding to a quasiparticle peak at the Fermi level, while the occupation imbalance vanishes, $\langle\hat{\Theta}\rangle=0$.
    \item[2)] In the paramagnetic Mott insulating phase (MI) the absence of quasiparticles is marked by $Z_{\vb{e},\vb{o}}=0$ and each site has the same average density, $\langle\hat{\Theta}\rangle=0$.
    \item[3)] In the density-wave phase (DW) the imbalance $\langle\hat{\Theta}\rangle$ is finite.~At quarter filling and for low temperatures the DW phase remains metallic \cite{pietig1999reentrant} while at half filling its single-particle spectrum is always gapped due to the checkerboard potential \cite{kapcia2017doping}, similarly to a band insulator.~The quasiparticle residue is finite in this phase, $Z_{\vb{e},\vb{o}}>0$.
\end{itemize}
For all calculations, we have introduced a finite initial value of the staggered potential usually corresponding to $\langle\hat \Theta\rangle\approx 0.2$ for starting the RDMFT iterations.~Furthermore, we use a lattice of size $N_s=32\times 32$ and set $t=1$, i.e.~the hopping amplitude is the unit of energy.

We have investigated both quarter filling ($\sum_{\bm{i}}\ev*{\hat{n}_{\bm{i}\up}}=\sum_{\bm{i}}\ev*{\hat{n}_{\bm{i}\down}}=N_s/4$) and half filling ($\sum_{\bm{i}}\ev*{\hat{n}_{\bm{i}\up}}=\sum_{\bm{i}}\ev*{\hat{n}_{\bm{i}\down}}=N_s/2$).~We analyzed the effects of temperature and cavity-mediated long-range interactions for fixed short-range interaction strengths and, at low temperature and half filling, determined the phase diagram as a function of $U_s$ and $U_l$.

\begin{figure}[!]
\includegraphics[width=1\linewidth]{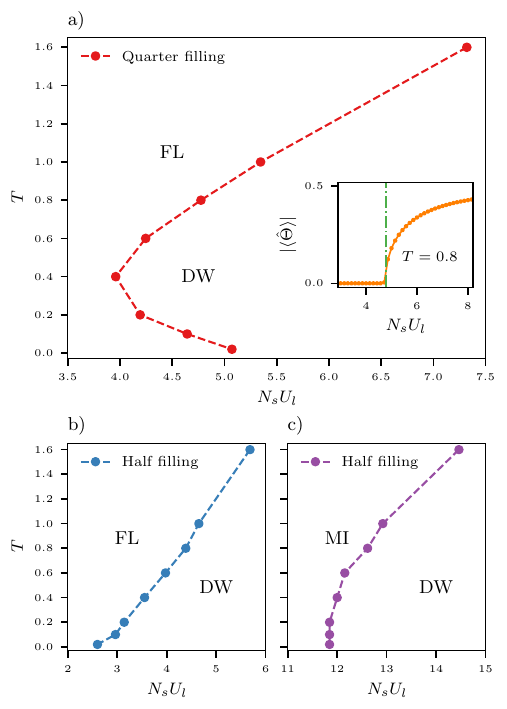}
\caption{\label{fig:qtfil} Phase diagram of the extended Fermi-Hubbard model with cavity-mediated long-range interactions at quarter and half filling in the $T$-$U_l$ plane.~The distinct phases are denoted as: FL (homogeneous Fermi liquid), MI (Mott insulator), DW (checkerboard density wave).~For both a) and c) the short-range interaction strength is fixed at $U_s=16$, while for b) its value is $U_s=3$.~At quarter filling a) we observe reentrant behavior at low temperatures.~In contrast, no such behavior is present at half filling, i.e.~in b) and c).~The inset in a) shows the $U_l$ dependence of the occupation imbalance $\langle\hat{\Theta}\rangle$ at temperature $T=0.8$ where the green dot-dashed line marks the FL-DW transition at $N_sU_l=4.77$.}
\end{figure}

In Fig.\,\ref{fig:qtfil}(a) the quarter filling $T$-$U_l$-phase diagram for $U_s=16$ is shown.~We found a transition from a homogeneous FL phase into a DW phase.~The curve for the critical long-range interaction strength as a function of temperature $U_l^c(T)$ shows a negative slope for $T\leq 0.4$.~This implies that in the interval $3.96<N_sU_l<5.07$, increasing temperature leads to a crystallization of the FL phase into a DW phase.~As we have neglected the possibility of magnetic ordering, at quarter filling the entropy of the spin degrees of freedom for each occupied site in the deep DW phase is approximately $\ln2$.~Hence, the entropy contribution reduces the free energy in the DW by a larger amount in contrast to the entropy of the FL state which is proportional to $T$ \cite{georges1996dynamical, pietig1999reentrant}, thus leading to the FL-DW transition with increasing temperature.~Further increasing $T$ leads to the expected melting of the DW phase and a reentrance into the homogeneous FL phase.~This is similar to earlier DMFT findings for the extended Hubbard model with nearest-neighbor interactions on a Bethe lattice \cite{pietig1999reentrant}.

Figures \ref{fig:qtfil}(b) and \ref{fig:qtfil}(c) display the corresponding phase diagram at half filling.~For $U_s=3$, a FL-DW transition is observed [Fig.\,\ref{fig:qtfil}(b)], while for $U_s=16$ the transition occurs between MI and DW states [Fig.\,\ref{fig:qtfil}(c)].~At half filling and large occupation imbalance, $\langle\hat{\Theta}\rangle\approx\pm1$, the DW phase consists of one nearly empty and one approximately doubly occupied sublattice, forming local singlets with no spin degeneracy and hence zero entropy.~In contrast, in the FL the entropy is proportional to $T$, and in the MI phase each site is singly occupied, resulting in an entropy of $\ln2$ per site.~Thus for both FL and MI the free energy is lowered relative to the one deep in the DW phase.~As a result, there is no entropic advantage to forming DW order upon heating, so $U_l^c(T)$ has a positive slope and no reentrant behavior occurs at half filling.~We point out that with increasing temperature the FL and MI phases are expected to undergo a crossover into a bad metal and a bad insulator state, respectively \cite{park2008cluster}.~However, the occupation imbalance $\langle\hat{\Theta}\rangle$ still serves as an order parameter for the transition into the DW state.

\begin{figure}[!]
\includegraphics[width=1\linewidth]{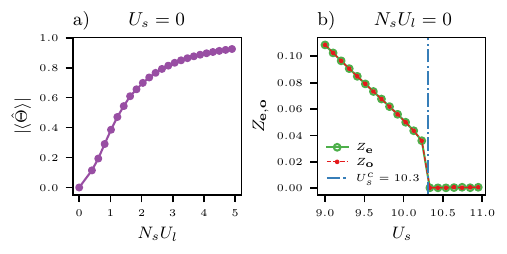}
\caption{\label{fig:nest} a) Occupation imbalance $\langle\hat{\Theta}\rangle$ for increasing long-range interaction strength $U_l$ at half filling, for $U_s=0$ and $T=0.02$.~Due to the perfect nesting of Fermi surface, the imbalance becomes finite as soon as the long-range interactions are non zero signaling the transition to the density-wave phase.~b) Quasiparticle weight $Z_{\vb{e},\vb{o}}$ as a function of the short-range interaction strength $U_s$ at half filling, for $U_l=0$ and $T=0.02$.~The paramagnetic Mott transition is characterized by a linear decrease of $Z_{\vb{e},\vb{o}}$ for increasing $U_s$ until it becomes zero at $U_s^c=10.3$ (dot-dashed green line) marking the vanishing of the Fermi liquid solution obtained using RDFMT.}
\end{figure}

At half filling and fixed temperature ($T=0.02$), we have investigated the phase transitions as a function of $U_l$ and $U_s$.~In the absence of Hubbard short-range interactions ($U_s=0$), the DW phase already emerges for vanishingly small long-range interaction strength [Fig.\,\ref{fig:nest}(a)].~This is due to the perfect nesting of the Fermi surface at half filling on a square lattice \cite{keeling2014fermionic,piazza2014umklapp,chen2014superradiance}.~As demonstrated in \cite{chen2014superradiance} for noninteracting spinless fermions, the critical pump-cavity coupling strength for the superradiant self-organization phase transition is inversely proportional to the fermionic density-density susceptibility.~For $U_s=U_l=0$, i.e.\,noninteracting spinful fermions on a half-filled square lattice, the density-density susceptibility shows a strong response for zero-energy excitations at the wave vector $\bm{q}=(\pi,\pi)$.~This response decreases for increasing $U_s$ \cite{hafermann2014collective} while keeping $U_l=0$.~In the commensurate configuration between cavity, pump and static square optical lattice ($k_p\approx k_c= k_{\rm latt}=k$), the cavity-mediated long-range interactions couple to the atomic density exactly at $\bm{q}$.~In particular, the sum in the second term of \eqref{eq5} can be written as 
\begin{equation}
    \sum_{\bm{i}\sigma}(-1)^{\abs{\bm{i}}}\hat{n}_{\bm{i}\sigma}=\sum_{\bm{i}\sigma}{\rm e}^{\ii\bm{q}\cdot\bm{i}}\hat{n}_{\bm{i}\sigma}.
\end{equation}
The strong density-density response of noninteracting fermions on a square lattice at half filling for zero energy excitations at the nesting vector $\bm{q}$ leads to the onset of the DW phase at arbitrarily small $U_l$.

\begin{figure}[!]
\includegraphics[width=1\linewidth]{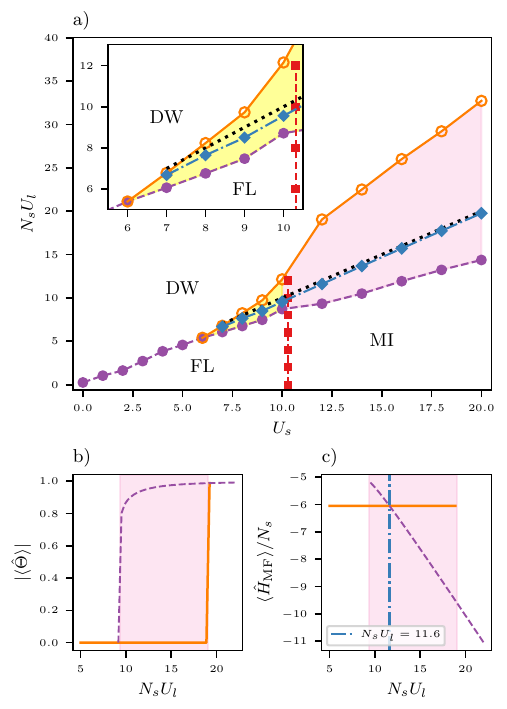}
\caption{\label{fig:pd} a) Phase diagram of the extended Fermi-Hubbard model with cavity-mediated long-range interactions at half filling and for $T=0.02$.~The distinct phases are denoted as FL (homogeneous Fermi liquid), MI (Mott insulator), DW (checkerboard density wave).~The FL–MI transition is marked by red squares.~The purple circle dots correspond to the border of the region of metastability of the DW solution obtained within RDMFT, while the orange empty circles represent the border of the region of metastability of the homogeneous FL and MI solutions.~The region where the MI and DW solutions coexist is depicted in shaded pink, and the region where the FL and DW solutions coexist is represented by the yellow shaded area.~The inset shows a zoom into the FL-DW coexistence region.~The diamond blue markers represent the thermodynamic phase transition obtained by comparing the energies of the homogeneous FL and MI solutions with the DW solution obtained within RDMFT.~These points follow closely the dotted black line $N_sU_l=U_s$ obtained by analyzing the model in the atomic limit.~Plots b) and c) show the hysteretic behavior of the occupation imbalance $\abs*{\langle\hat{\Theta}\rangle}$ and the expectation value of the mean-field Hamiltonian $\ev*{\hat{H}_{\rm MF}}$ [Eq.\,\eqref{efH}] for $U_s=12$ and $T=0.02$, respectively.~In b) and c) the dashed purple line indicates the DW solution while the continuous orange line corresponds to the MI solution.~The thermodynamic phase transition is indicated in c) by the vertical dot-dashed blue line at $N_sU_l=11.6$.}
\end{figure}

The full phase diagram at half filling is shown in Fig.\,\ref{fig:pd}(a).~For small short-range interactions, increasing $U_l$ leads to a FL-DW transition.~In this region of small $U_s$ and $U_l$, the FL-DW phase boundary remains close to the line $N_sU_l=U_s$.~For small $U_l$, increasing the Hubbard short-range interactions $U_s$ from intermediate to strong, leads to a FL-MI transition, where the critical value $U_s^c$ remains constant for increasing $U_l$ (see red square markers).~This is because in this regime the mean-field Hamiltonian \eqref{efH} reduces to the standard Fermi-Hubbard model due the vanishing imbalance, $\langle\hat{\Theta}\rangle=0$.~The critical Hubbard interaction for the FL-MI transition, obtained from the vanishing of $Z_{\vb{e},\vb{o}}$ depicted in Fig.\,\ref{fig:nest}(b), remains at $U_s^c= 10.3$ independent of $U_l$ (red square markers).~We point out that at low temperatures, DMFT calculations predict a region across the paramagnetic Mott transition where both the FL and the MI solution coexist \cite{georges1996dynamical,park2008cluster}.~At zero temperature and for $U_l=0$, the thermodynamic FL-MI phase transition occurs precisely at the critical Hubbard interaction $U_s^c$ which is determined by the breakdown of the FL solution obtained using DMFT.~We therefore use this criterion to determine the FL–MI transition within RDMFT.

For sufficiently large $U_s$, increasing the long-range interaction $U_l$ drives a MI-DW transition.~The expected thermodynamic phase transition can be understood by analyzing the model in atomic limit of $t=0$ at zero temperature \cite{panas2017spectral,hruby2018metastability}.~In this limit the eigenstates of the Hamiltonian \eqref{eq5} are Fock states with well-defined number of particles at each site.

Assuming half filling and a given finite occupation imbalance $\langle\hat{\Theta}\rangle>0$, the competition between $U_s$ and $U_l$ leads to the formation of domains, where a fraction $\lvert\langle\hat{\Theta}\rangle\rvert$ of the sites is in the DW domain while the remaining fraction is in the MI domain.~The energy per particle (per site) of such a state has the following form (see appendix A)
\begin{equation}\label{eq12}
    \frac{\ev*{\hat H}}{N_s}=\frac{U_s}{2}\abs*{\langle\hat{\Theta}\rangle}-\frac{N_sU_l}{2}\langle\hat{\Theta}\rangle^2-\tilde{\mu}.
\end{equation}
In the $T\rightarrow 0$ limit, the grand potential coincides with $\ev*{\hat H}$.~For weak long-range interaction strength, $N_sU_l/U_s\ll1$, Eq.\,\eqref{eq12} has a single minimum at $\langle\hat{\Theta}\rangle=0$ corresponding to a MI state, whereas for $N_sU_l/U_s>1$, two degenerate minima at $\langle\hat{\Theta}\rangle=\pm1$ emerge, corresponding to a DW phase.~In the vicinity of $N_sU_l/U_s\approx 1$, these minima at $\langle\hat{\Theta}\rangle=0$ and at $\langle\hat{\Theta}\rangle=\pm1$ become close in energy (see appendix A), indicating a first-order transition with competing local minima.~A similar atomic-limit argument was used in \cite{panas2017spectral} where lattice bosons coupled to a transversely pumped cavity were studied with bosonic DMFT, where a region of coexisting MI and DW solutions was found.~Subsequently, the metastability of these two states was observed experimentally in \cite{hruby2018metastability} in the bosonic case.

This first-order character is directly reflected in our RDMFT calculations at intermediate to strong $U_s$.~Near $N_sU_l/U_s\approx 1$, we observe that the converged solution of the algorithm depends significantly on the initialization of the self-consistency loop.~For strong (intermediate) 
$U_s$ values, we find that initializing iterations from the MI (FL) phase and increasing $U_l$ leads to a larger critical value $U_l^c$ for the MI-DW (FL-DW) transition [orange empty circles in Fig.\,\ref{fig:pd}(a)] than starting from the DW phase and decreasing $U_l$ for a fixed $U_s$ [purple circles in Fig.\,\ref{fig:pd}(a)], resulting in the hysteresis of $\langle\hat{\Theta}\rangle$ [Fig.\,\ref{fig:pd}(b)]. 

As summarized in Fig.\,\ref{fig:pd}(a), this leads to well-defined regions in which either the FL or the MI solutions coexist with the DW solution.~For moderate values of $U_s$ and $U_l$, we find an FL-DW coexistence region (yellow shaded area) while for strong $U_s$ and $U_l$ we find a MI-DW coexistence region (light-pink shaded area).~A similar result was found in \cite{kapcia2017doping} for the extended Hubbard model with nearest neighbor interactions on a Bethe lattice.

To calculate the thermodynamic transition, we follow the approach used in \cite{panas2017spectral}, approximating the grand potential by the expectation value of the effective mean-field Hamiltonian \eqref{efH}, $\Omega\approx\ev*{\hat{H}_{\rm MF}}$, and neglecting the contribution $-TS$, where $S$ is the entropy.~While this approximation is exact only at $T\rightarrow 0$, for the temperature $T=0.02$ used in Fig.\,\ref{fig:pd}, the maximum value of the term $-TS$ contributing to the grand potential is less than $0.3$\% of the energy and has negligible influence on the prediction of the thermodynamic phase transition.~In RDMFT, the expectation values of the local terms in the Hamiltonian \eqref{efH} are obtained directly from the exact diagonalization of the effective impurity Hamiltonian \eqref{eq:H_AIM}.~In appendix B we demonstrate how the kinetic energy corresponding to the expectation value of the tunneling term in \eqref{efH} can be calculated from local quantities in RDMFT.

As shown in Fig.\,\ref{fig:pd}(c), where $U_s=12$, the energy of DW solution decreases linearly with $U_l$ in the coexistence region and becomes lower than the energy of the MI solution at $U_l=11.6$, indicating the thermodynamic MI-DW phase transition.~The resulting phase transition points [blue diamonds in Fig.\,\ref{fig:pd}(a)] are in good agreement with the line $N_sU_l=U_s$ (dotted black line) obtained from the atomic limit analysis of the model.~For the system sizes investigated ($8\times8$, $16\times16$, $32\times32$), no significant finite-size effects were observed for this phase boundary as well as for the boundaries of the coexistence region.

At low temperatures and finite Hubbard interactions,~DMFT-based approaches to the standard Fermi–Hubbard model predict the emergence of antiferromagnetic solutions at half filling due to superexchange couplings \cite{georges1996dynamical}.~In one and two dimensions, however, the Hohenberg–Mermin–Wagner (HMW) theorem forbids true long-range magnetic order at finite temperature in the thermodynamic limit \cite{mermin1966absence,hohenberg1967existence}.~More recent studies indicate that in finite or disordered two dimensional systems, magnetic order remains nonvanishing below a sizable $T_c$ in spite of the strong infrared fluctuations \cite{palle2021physical}.~The present low-temperature results therefore provide a meaningful reference description and are expected to be relevant in regimes where magnetic order is suppressed or frustrated.~A detailed investigation of the emergence of magnetic order and its competition with cavity-induced density order remains an important direction for future work.

\section{\label{sec:conc}Conclusion}

We have investigated the phases of the extended Fermi-Hubbard model with cavity-mediated long-range interactions on a square lattice, both at half and quarter filling, using RDMFT.~We included the long-range interactions via a static mean-field decoupling.~At quarter filling, we observed reentrant behavior in a region of the phase diagram where the systems undergoes a phase transition from the homogeneous Fermi-liquid phase to the density-wave phase and back to the homogeneous Fermi-liquid phase as temperature is increased.~At half filling, we found that for small Hubbard- and long-range interactions a Fermi-liquid phase exists.~However, for vanishing Hubbard $U_s$ the instability of this phase towards a checkerboard density-wave insulating phase occurs for arbitrarily small long-range interaction strengths due to the perfect nesting of the Fermi surface at half filling \cite{keeling2014fermionic,piazza2014umklapp,chen2014superradiance}.~For intermediate and strong Hubbard $U_s$ and long-range interactions $U_l$, we observe a coexistence region in the phase diagram where the homogeneous Fermi-liquid, Mott insulating, and density wave solutions of RDMFT are metastable.~We observed that the occupation imbalance shows hysteretic behavior in the coexistence region, which is characteristic of a first-order phase transition.~By comparing the energy of the different RDMFT solutions in that region, we determined the thermodynamic phase boundary for the transition into the density-wave phase.

\begin{acknowledgments}
We thank Lo\"ic Philoxene for comments that greatly improved the present work.~This work was funded by the Deutsche Forschungsgemeinschaft (DFG, German Research Foundation) under Project No.\,$557989325$ as an individual grant (Sachbeihilfe) with the project ID HO 2407/12-1.~The authors gratefully acknowledge the computing time provided to them at the NHR Center NHR@SW at Goethe University Frankfurt.~This is funded by the Federal Ministry of Education and Research, and the state governments participating on the basis of the resolutions of the GWK for national high performance computing at universities \cite{nhr}.~The authors gratefully acknowledge the Gauss Centre for Supercomputing e.V.\,\cite{gauss} for funding this project by providing computing time through the John von Neumann Institute for Computing (NIC) on the GCS
Supercomputer JUWELS at Jülich Supercomputing Centre (JSC).
\end{acknowledgments}

\appendix

\section{Metastability in the atomic limit}

\begin{figure}[h]
\includegraphics[width=1\linewidth]{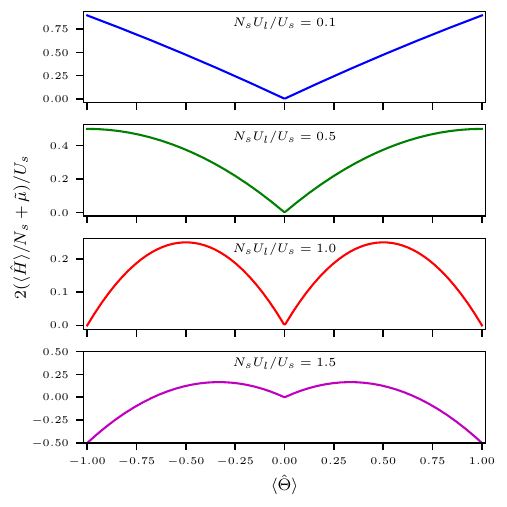}
\caption{\label{fig:fig5}  Expectation value of the energy in the atomic limit \eqref{eqA1} for different values of the ratio $N_sU_l/U_s$.}
\end{figure}

In this appendix we present a derivation justifying the metastability found in our results by analyzing the expectation value for the effective Hamiltonian \eqref{eq5} in the atomic limit, following \cite{panas2017spectral,hruby2018metastability}.~In this limit we have $t=0$, and the effective Hamiltonian \eqref{eq5} assumes the following form
\begin{equation}\label{eqA1}
    \hat{H}=U_s\sum_{\bm{i}}\hat{n}_{\bm{i}\up}\hat{n}_{\bm{i}\down}-\frac{U_l}{2}\Big(\sum_{\bm{i}\sigma}(-1)^{\abs{\bm{i}}}\hat{n}_{\bm{i}\sigma}\Big)^2- \sum_{\bm{i}\sigma}\tilde{\mu}_\sigma\hat{n}_{\bm{i}\sigma}.
\end{equation}
At half filling for a balanced configuration between the spins, we have that $\tilde{\mu}_\up=\tilde{\mu}_\down\equiv\tilde{\mu}$ and $\sum_{\bm{i}}\ev*{\hat{n}_{\bm{i}\up}}=\sum_{\bm{i}}\ev*{\hat{n}_{\bm{i}\down}}=N_s/2$.~In this case, the interplay between $U_s$ and $U_l$ creates a state where most lattice sites are singly occupied, with a subset of the sites on one sublattice doubly occupied, while its complement on the other sublattice remains empty, such that the total filling is conserved.~Assuming that the doubly occupied sites reside on the even sublattice, the total particle numbers on the even and odd sublattices are $N_{e}=N_s/2+N_d$ and $N_{o}=N_s/2-N_d$, respectively, where $N_d$ is the number of doubly occupied sites.~The resulting average occupation imbalance is $\ev*{\hat{\Theta}}=(N_{e}-N_{o})/N_s=2N_d/N_s$.~Accounting for the equivalent configuration in which the roles of the sublattices are exchanged, we have that $N_d=|\langle\hat{\Theta}\rangle|N_s/2$.~These doubly occupied sites increase the system's energy due to the onsite Hubbard interaction $U_s$, so that the expectation value of the short-range interaction term becomes
\begin{equation}
   E_{\rm int}=U_s\sum_{\bm{i}}\ev*{\hat{n}_{\bm{i}\up}\hat{n}_{\bm{i}\down}}=U_sN_d=\frac{1}{2}U_s|\langle\hat{\Theta}\rangle|N_s. 
\end{equation}
Thus, the expectation value of the Hamiltonian \eqref{eqA1} can be written as,
\begin{equation}\label{eqA3}
    \frac{\ev*{\hat H}}{N_s}=\frac{U_s}{2}\abs*{\langle\hat{\Theta}\rangle}-\frac{N_sU_l}{2}\langle\hat{\Theta}\rangle^2-\tilde{\mu},
\end{equation}
where we have assumed a vanishing variance of the occupation imbalance such that $\ev*{\hat{\Theta}^2}=\langle\hat{\Theta}\rangle^2$.~In Fig.\,\ref{fig:fig5} we plot \eqref{eqA3} for different values of the ratio $N_sU_l/U_s$.~As can be observed, for $N_sU_l/U_s<1$, the atomic limit energy has a minimum at $\langle\hat{\Theta}\rangle=0$ corresponding to the MI phase.~At $N_sU_l/U_s=1$, the expectation value \eqref{eqA3} shows three minima, one at $\langle\hat{\Theta}\rangle=0$ and the other two at $\langle\hat{\Theta}\rangle=\pm1$.~For $N_sU_l/U_s>1$, the minima at $\langle\hat{\Theta}\rangle=\pm1$ become lower in energy than the local minimum at $\langle\hat{\Theta}\rangle=0$ leading to the stabilization of a DW phase.

\section{Kinetic Energy in RDMFT}
Apart from the kinetic energy term contribution to the expectation value of the mean-field Hamiltonian
\begin{equation}
    \begin{aligned}
    \ev*{\hat{H}_{\rm MF}}=&-\sum_{\bm{ij}\sigma}t_{\bm{ij}}\ev*{\hat{c}^\dagger_{\bm{i}\sigma}\hat{c}^{\vphantom\dagger}_{\bm{j}\sigma}}+U_s\sum_{\bm{i}}\ev*{\hat{n}_{\bm{i}\up}\hat{n}_{\bm{i}\down}}
    \\
    &-\sum_{\bm{i}\sigma}\tilde{\mu}_\sigma\ev{\hat{n}_{\bm{i}\sigma}}-N_s^2\frac{U_l}{2}\langle\hat{\Theta}\rangle^2
    \end{aligned}
\end{equation}
all other terms are local and can be computed directly from the impurity solver within RDMFT.~Following \cite{geissler2018lattice}, here we show how the expectation value of the kinetic energy can be expressed in terms of local quantities and calculated using RDMFT. 

We first point out that, in terms of the imaginary time lattice Green's function, the expectation value of the operators in the hopping term of the mean field Hamiltonian \eqref{efH} can be written as
\begin{equation}
    \begin{aligned}
        \ev*{\hat{c}^\dagger_{\bm{i}\sigma}\hat{c}^{\vphantom\dagger}_{\bm{j}\sigma}}=&\lim_{\eta\to 0^+}[\vb{G}_\sigma(\tau=-\eta)]_{\bm{ji}} 
        \\
        =&\frac{1}{\beta}\lim_{\eta\to 0^+}\sum_m[\vb{G}_\sigma(\ii\omega_m)]_{\bm{ji}}{\rm e}^{\ii\omega_m\eta}.
    \end{aligned}
\end{equation}
Thus, the kinetic energy becomes
\begin{equation}
    \begin{aligned}
        E_{\rm kin}=&-\sum_{\bm{ij}\sigma}t_{\bm{ij}}\ev*{\hat{c}^\dagger_{\bm{i}\sigma}\hat{c}^{\vphantom\dagger}_{\bm{j}\sigma}}
        \\
        =&-\frac{1}{\beta}\lim_{\eta\to 0^+}\sum_m\sum_{\bm{ij}\sigma}t_{\bm{ij}}[\vb{G}_\sigma(\ii\omega_m)]_{\bm{ji}}{\rm e}^{\ii\omega_m\eta}.
    \end{aligned}
\end{equation}
Making use of the lattice Dyson equation $[\vb{G}_\sigma(\ii\omega_m)]^{-1}_{\bm{ij}}=[\tilde{\mu}_\sigma+\ii\omega_m+V_{\bm{i}}-\Sigma_{\bm{i},\sigma}(\ii\omega_m)]\delta_{\bm{ij}}+t_{\bm{ij}}$, we note that
\begin{equation}\label{eqB4}
    \begin{aligned}
        &\sum_{\bm{j}}[\vb{G}_\sigma(\ii\omega_m)]^{-1}_{\bm{ij}}[\vb{G}_\sigma(\ii\omega_m)]_{\bm{ji}}=1
        \\
        &\sum_{\bm{j}}\{[\tilde{\mu}_\sigma+\ii\omega_m+V_{\bm{i}}-\Sigma_{\bm{i},\sigma}(\ii\omega_m)]\delta_{\bm{ij}}+t_{\bm{ij}}\}
        \\
        &\qquad\times[\vb{G}_\sigma(\ii\omega_m)]_{\bm{ji}}=1.
    \end{aligned}
\end{equation}
From the equation above, using the local Dyson equation $\Sigma_{\bm{i},\sigma}(\ii\omega_m)=\mathcal{G}^{(\bm{i})}_\sigma(\ii\omega_m)^{-1}-G_{\sigma}^{(\bm{i})}(\ii\omega_m)^{-1}$ and the relation $G_{\sigma}^{(\bm{i})}(\ii\omega_m)=[\vb{G}_\sigma(\ii\omega_m)]_{\bm{ii}}$, it can be demonstrated that
\begin{equation}
\begin{aligned}
        -\sum_{\bm{j}}t_{\bm{ij}}[\vb{G}_\sigma(\ii\omega_m)]_{\bm{ji}}=&[\tilde{\mu}_\sigma+\ii\omega_m+V_{\bm{i}}-\mathcal{G}^{(\bm{i})}_\sigma(\ii\omega_m)^{-1}]
        \\
        &\times G_{\sigma}^{(\bm{i})}(\ii\omega_m).
\end{aligned}
\end{equation}
Notice that the $+1$ term on the right-hand side of Eq.\,\eqref{eqB4} cancels with the one coming from the multiplication of the local Green's function by its inverse $G_{\sigma}^{(\bm{i})}(\ii\omega_m)^{-1}G_{\sigma}^{(\bm{i})}(\ii\omega_m)=1$ on the left-hand side which appears after making the substitution of the self-energy using the local Dyson equation. 
Hence, the kinetic energy can be finally written as
\begin{equation}
\begin{aligned}
        E_{\rm kin}=&\frac{2}{\beta}\lim_{\eta\to 0^+}\sum_{m\geq0}{\rm e}^{\ii\omega_m\eta}
        \\
        &\times\sum_{\bm{i}\sigma}[\tilde{\mu}_\sigma+\ii\omega_m+V_{\bm{i}}-\mathcal{G}^{(\bm{i})}_\sigma(\ii\omega_m)^{-1}]G_{\sigma}^{(\bm{i})}(\ii\omega_m).
\end{aligned}
\end{equation}
Numerically, only a finite number of positive Matsubara frequencies is considered in the calculation which implies that the equal time limit is not obtained by simply imposing $\eta=0$.~The solution is found by finding for every site the value of $\eta$ satisfying the relation
\begin{equation}
    \frac{2}{\beta}\lim_{\eta\to 0^+}\sum_{m\geq0}{\rm e}^{\ii\omega_m\eta}G_{\sigma}^{(\bm{i})}(\ii\omega_m)=\ev{\hat{n}_{\bm{i}\sigma}}_{\rm AIM},
\end{equation}
where $\ev{\hat{n}_{\bm{i}\sigma}}_{\rm AIM}$ is obtained by solving the Anderson impurity Hamiltonian \eqref{eq:H_AIM}.

\begin{figure}[h]
\includegraphics[width=\linewidth]{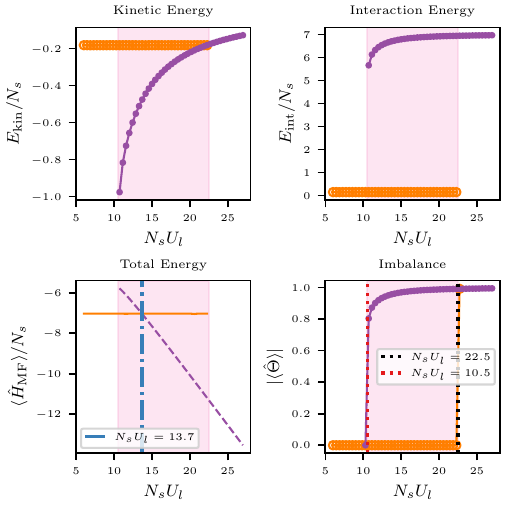}
\caption{\label{fig:fig6}Kinetic energy, short-range interaction energy, total energy, and occupation imbalance calculated using RDMFT across the MI-DW phase transition for $U_s=14$ and $T=0.02$.~The DW solution is shown in purple and the MI solution is shown in orange.}
\end{figure}

In Fig.\,\ref{fig:fig6}, we show the energy calculated across the MI-DW phase transition for the different MI and DW  solutions obtained within RDMFT.
\par\noindent
\par\noindent
\clearpage
\bibliography{apssamp}

@article{landig2016quantum,
  title={Quantum phases from competing short-and long-range interactions in an optical lattice},
  author={Landig, Renate and Hruby, Lorenz and Dogra, Nishant and Landini, Manuele and Mottl, Rafael and Donner, Tobias and Esslinger, Tilman},
  journal={Nature},
  volume={532},
  number={7600},
  pages={476--479},
  year={2016},
  publisher={Nature Publishing Group UK London}
}

@article{zhang2021observation,
  title={Observation of a superradiant quantum phase transition in an intracavity degenerate Fermi gas},
  author={Zhang, Xiaotian and Chen, Yu and Wu, Zemao and Wang, Juan and Fan, Jijie and Deng, Shujin and Wu, Haibin},
  journal={Science},
  volume={373},
  number={6561},
  pages={1359--1362},
  year={2021},
  publisher={American Association for the Advancement of Science}
}

@article{roux2020strongly,
  title={Strongly correlated Fermions strongly coupled to light},
  author={Roux, Kevin and Konishi, Hideki and Helson, Victor and Brantut, Jean-Philippe},
  journal={Nature Communications},
  volume={11},
  number={1},
  pages={2974},
  year={2020},
  publisher={Nature Publishing Group UK London}
}

@article{georges1996dynamical,
  title={Dynamical mean-field theory of strongly correlated fermion systems and the limit of infinite dimensions},
  author={Georges, Antoine and Kotliar, Gabriel and Krauth, Werner and Rozenberg, Marcelo J},
  journal={Reviews of Modern Physics},
  volume={68},
  number={1},
  pages={13},
  year={1996},
  publisher={APS}
}

@article{snoek2008antiferromagnetic,
  title={Antiferromagnetic order of strongly interacting fermions in a trap: real-space dynamical mean-field analysis},
  author={Snoek, M and Titvinidze, I and T{\H{o}}ke, C and Byczuk, Krzysztof and Hofstetter, Walter},
  journal={New Journal of Physics},
  volume={10},
  number={9},
  pages={093008},
  year={2008},
  publisher={IOP Publishing}
}

@article{pietig1999reentrant,
  title={Reentrant charge order transition in the extended Hubbard model},
  author={Pietig, R and Bulla, Ralf and Blawid, S},
  journal={Physical Review Letters},
  volume={82},
  number={20},
  pages={4046},
  year={1999},
  publisher={APS}
}

@article{keeling2014fermionic,
  title={Fermionic superradiance in a transversely pumped optical cavity},
  author={Keeling, J and Bhaseen, MJ and Simons, BD},
  journal={Physical Review Letters},
  volume={112},
  number={14},
  pages={143002},
  year={2014},
  publisher={APS}
}

@article{piazza2014umklapp,
  title={Umklapp superradiance with a collisionless quantum degenerate fermi gas},
  author={Piazza, Francesco and Strack, Philipp},
  journal={Physical Review Letters},
  volume={112},
  number={14},
  pages={143003},
  year={2014},
  publisher={APS}
}

@article{chen2014superradiance,
  title={Superradiance of degenerate Fermi gases in a cavity},
  author={Chen, Yu and Yu, Zhenhua and Zhai, Hui},
  journal={Physical Review Letters},
  volume={112},
  number={14},
  pages={143004},
  year={2014},
  publisher={APS}
}

@article{hruby2018metastability,
  title={Metastability and avalanche dynamics in strongly correlated gases with long-range interactions},
  author={Hruby, Lorenz and Dogra, Nishant and Landini, Manuele and Donner, Tobias and Esslinger, Tilman},
  journal={Proceedings of the National Academy of Sciences},
  volume={115},
  number={13},
  pages={3279--3284},
  year={2018},
  publisher={National Academy of Sciences}
}

@article{maschler2005cold,
  title={Cold atom dynamics in a quantum optical lattice potential},
  author={Maschler, Christoph and Ritsch, Helmut},
  journal={Physical Review Letters},
  volume={95},
  number={26},
  pages={260401},
  year={2005},
  publisher={APS}
}

@article{maschler2008ultracold,
  title={Ultracold atoms in optical lattices generated by quantized light fields},
  author={Maschler, Christoph and Mekhov, Igor B and Ritsch, Helmut},
  journal={The European Physical Journal D},
  volume={46},
  pages={545--560},
  year={2008},
  publisher={Springer}
}

@article{dogra2016phase,
  title={Phase transitions in a Bose-Hubbard model with cavity-mediated global-range interactions},
  author={Dogra, Nishant and Brennecke, Ferdinand and Huber, Sebastian D and Donner, Tobias},
  journal={Physical Review A},
  volume={94},
  number={2},
  pages={023632},
  year={2016},
  publisher={APS}
}

@article{niederle2016ultracold,
  title={Ultracold bosons with cavity-mediated long-range interactions: A local mean-field analysis of the phase diagram},
  author={Niederle, Astrid E and Morigi, Giovanna and Rieger, Heiko},
  journal={Physical Review A},
  volume={94},
  number={3},
  pages={033607},
  year={2016},
  publisher={APS}
}

@article{panas2017spectral,
  title={Spectral properties and phase diagram of correlated lattice bosons in an optical cavity within bosonic dynamical mean-field theory},
  author={Panas, Jaromir and Kauch, Anna and Byczuk, Krzysztof},
  journal={Physical Review B},
  volume={95},
  number={11},
  pages={115105},
  year={2017},
  publisher={APS}
}

@article{griesmaier2005bose,
  title={Bose-Einstein condensation of chromium},
  author={Griesmaier, Axel and Werner, J{\"o}rg and Hensler, Sven and Stuhler, J{\"u}rgen and Pfau, Tilman},
  journal={Physical Review Letters},
  volume={94},
  number={16},
  pages={160401},
  year={2005},
  publisher={APS}
}

@article{baumann2010dicke,
  title={Dicke quantum phase transition with a superfluid gas in an optical cavity},
  author={Baumann, Kristian and Guerlin, Christine and Brennecke, Ferdinand and Esslinger, Tilman},
  journal={Nature},
  volume={464},
  number={7293},
  pages={1301--1306},
  year={2010},
  publisher={Nature Publishing Group UK London}
}

@article{micnas1988superconductivity,
  title={Superconductivity in a narrow-band system with intersite electron pairing in two dimensions: A mean-field study},
  author={Micnas, R and Ranninger, J and Robaszkiewicz, S and Tabor, S},
  journal={Physical Review B},
  volume={37},
  number={16},
  pages={9410},
  year={1988},
  publisher={APS}
}

@article{helson2023density,
  title={Density-wave ordering in a unitary Fermi gas with photon-mediated interactions},
  author={Helson, Victor and Zwettler, Timo and Mivehvar, Farokh and Colella, Elvia and Roux, Kevin and Konishi, Hideki and Ritsch, Helmut and Brantut, Jean-Philippe},
  journal={Nature},
  volume={618},
  number={7966},
  pages={716--720},
  year={2023},
  publisher={Nature Publishing Group UK London}
}

@article{piazza2014quantum,
  title={Quantum kinetics of ultracold fermions coupled to an optical resonator},
  author={Piazza, Francesco and Strack, Philipp},
  journal={Physical Review A},
  volume={90},
  number={4},
  pages={043823},
  year={2014},
  publisher={APS}
}

@article{mivehvar2021cavity,
  title={Cavity QED with quantum gases: new paradigms in many-body physics},
  author={Mivehvar, Farokh and Piazza, Francesco and Donner, Tobias and Ritsch, Helmut},
  journal={Advances in Physics},
  volume={70},
  number={1},
  pages={1--153},
  year={2021},
  publisher={Taylor \& Francis}
}

@article{sotnikov2015critical,
  title={Critical entropies and magnetic-phase-diagram analysis of ultracold three-component fermionic mixtures in optical lattices},
  author={Sotnikov, Andrii},
  journal={Physical Review A},
  volume={92},
  number={2},
  pages={023633},
  year={2015},
  publisher={APS}
}

@article{park2008cluster,
  title={Cluster dynamical mean field theory of the Mott transition},
  author={Park, Hyowon and Haule, Kristjan and Kotliar, Gabriel},
  journal={Physical Review Letters},
  volume={101},
  number={18},
  pages={186403},
  year={2008},
  publisher={APS}
}

@article{bernien2017probing,
  title={Probing many-body dynamics on a 51-atom quantum simulator},
  author={Bernien, Hannes and Schwartz, Sylvain and Keesling, Alexander and Levine, Harry and Omran, Ahmed and Pichler, Hannes and Choi, Soonwon and Zibrov, Alexander S and Endres, Manuel and Greiner, Markus and others},
  journal={Nature},
  volume={551},
  number={7682},
  pages={579--584},
  year={2017},
  publisher={Nature Publishing Group UK London}
}

@article{labuhn2016tunable,
  title={Tunable two-dimensional arrays of single Rydberg atoms for realizing quantum Ising models},
  author={Labuhn, Henning and Barredo, Daniel and Ravets, Sylvain and De L{\'e}s{\'e}leuc, Sylvain and Macr{\`\i}, Tommaso and Lahaye, Thierry and Browaeys, Antoine},
  journal={Nature},
  volume={534},
  number={7609},
  pages={667--670},
  year={2016},
  publisher={Nature Publishing Group UK London}
}

@article{chicireanu2006simultaneous,
  title={Simultaneous magneto-optical trapping of bosonic and fermionic chromium atoms},
  author={Chicireanu, R and Pouderous, A and Barb{\'e}, R and Laburthe-Tolra, B and Mar{\'e}chal, E and Vernac, L and Keller, J-C and Gorceix, O},
  journal={Physical Review A},
  volume={73},
  number={5},
  pages={053406},
  year={2006},
  publisher={APS}
}

@article{lahaye2009physics,
  title={The physics of dipolar bosonic quantum gases},
  author={Lahaye, Thierry and Menotti, C and Santos, L and Lewenstein, M and Pfau, T},
  journal={Reports on Progress in Physics},
  volume={72},
  number={12},
  pages={126401},
  year={2009},
  publisher={IOP Publishing}
}

@article{zeiher2016many,
  title={Many-body interferometry of a Rydberg-dressed spin lattice},
  author={Zeiher, Johannes and Van Bijnen, Rick and Schau{\ss}, Peter and Hild, Sebastian and Choi, Jae-yoon and Pohl, Thomas and Bloch, Immanuel and Gross, Christian},
  journal={Nature Physics},
  volume={12},
  number={12},
  pages={1095--1099},
  year={2016},
  publisher={Nature Publishing Group UK London}
}

@article{baranov2012condensed,
  title={Condensed matter theory of dipolar quantum gases},
  author={Baranov, Mikhail A and Dalmonte, Marcello and Pupillo, Guido and Zoller, Peter},
  journal={Chemical Reviews},
  volume={112},
  number={9},
  pages={5012--5061},
  year={2012},
  publisher={ACS Publications}
}

@article{dagotto1994superconductivity,
  title={Superconductivity near phase separation in models of correlated electrons},
  author={Dagotto, E and Riera, J and Chen, YC and Moreo, A and Nazarenko, A and Alcaraz, F and Ortolani, F},
  journal={Physical Review B},
  volume={49},
  number={5},
  pages={3548},
  year={1994},
  publisher={APS}
}

@article{chattopadhyay1997phase,
  title={Phase diagram of the half-filled extended Hubbard model in two dimensions},
  author={Chattopadhyay, Biplab and Gaitonde, DM},
  journal={Physical Review B},
  volume={55},
  number={23},
  pages={15364},
  year={1997},
  publisher={APS}
}

@article{rosciszewski2003charge,
  title={Charge order in the extended Hubbard model},
  author={Ro{\'s}ciszewski, Krzysztof and Ole{\'s}, Andrzej M},
  journal={Journal of Physics: Condensed Matter},
  volume={15},
  number={49},
  pages={8363},
  year={2003},
  publisher={IOP Publishing}
}

@article{kapcia2017doping,
  title={Doping-driven metal-insulator transitions and charge orderings in the extended Hubbard model},
  author={Kapcia, KJ and Robaszkiewicz, S and Capone, Massimo and Amaricci, Adriano},
  journal={Physical Review B},
  volume={95},
  number={12},
  pages={125112},
  year={2017},
  publisher={APS}
}

@article{terletska2017charge,
  title={Charge ordering and correlation effects in the extended Hubbard model},
  author={Terletska, Hanna and Chen, Tianran and Gull, Emanuel},
  journal={Physical Review B},
  volume={95},
  number={11},
  pages={115149},
  year={2017},
  publisher={APS}
}

@article{terletska2018charge,
  title={Charge ordering and nonlocal correlations in the doped extended Hubbard model},
  author={Terletska, Hanna and Chen, Tianran and Paki, Joseph and Gull, Emanuel},
  journal={Physical Review B},
  volume={97},
  number={11},
  pages={115117},
  year={2018},
  publisher={APS}
}

@article{calandra2002metal,
  title={Metal-insulator transition and charge ordering in the extended Hubbard model at one-quarter filling},
  author={Calandra, M and Merino, J and McKenzie, Ross H},
  journal={Physical Review B},
  volume={66},
  number={19},
  pages={195102},
  year={2002},
  publisher={APS}
}

@article{piazza2013bose,
  title={Bose--Einstein condensation versus Dicke--Hepp--Lieb transition in an optical cavity},
  author={Piazza, Francesco and Strack, Philipp and Zwerger, Wilhelm},
  journal={Annals of Physics},
  volume={339},
  pages={135--159},
  year={2013},
  publisher={Elsevier}
}

@article{hafermann2014collective,
  title={Collective charge excitations of strongly correlated electrons, vertex corrections, and gauge invariance},
  author={Hafermann, Hartmut and van Loon, Erik GCP and Katsnelson, Mikhail I and Lichtenstein, Alexander I and Parcollet, Olivier},
  journal={Physical Review B},
  volume={90},
  number={23},
  pages={235105},
  year={2014},
  publisher={APS}
}

@article{hofstetter2018quantum,
  title={Quantum simulation of strongly correlated condensed matter systems},
  author={Hofstetter, Walter and Qin, Tao},
  journal={Journal of Physics B: Atomic, Molecular and Optical Physics},
  volume={51},
  number={8},
  pages={082001},
  year={2018},
  publisher={IOP Publishing}
}

@phdthesis{geissler2018lattice,
  title={Lattice-supersolids in bosonic quantum gases with Rydberg excitations},
  author={Gei{\ss}ler, Andreas},
  year={2018},
  school={Universit{\"a}tsbibliothek Johann Christian Senckenberg}
}

@article{merino2007nonlocal,
  title={Nonlocal coulomb correlations in metals close to a charge order insulator transition},
  author={Merino, Jaime},
  journal={Physical Review Letters},
  volume={99},
  number={3},
  pages={036404},
  year={2007},
  publisher={APS}
}

@article{merino2001superconductivity,
  title={Superconductivity mediated by charge fluctuations in layered molecular crystals},
  author={Merino, Jaime and McKenzie, Ross H},
  journal={Physical Review Letters},
  volume={87},
  number={23},
  pages={237002},
  year={2001},
  publisher={APS}
}

@article{mckenzie2001charge,
  title={Charge ordering and antiferromagnetic exchange in layered molecular crystals of the $\theta$ type},
  author={McKenzie, Ross H and Merino, J and Marston, JB and Sushkov, OP},
  journal={Physical Review B},
  volume={64},
  number={8},
  pages={085109},
  year={2001},
  publisher={APS}
}

@article{PhysRevA.87.051604,
  title = {Lattice-supersolid phase of strongly correlated bosons in an optical cavity},
  author = {Li, Yongqiang and He, Liang and Hofstetter, Walter},
  journal = {Physical Review A},
  volume = {87},
  issue = {5},
  pages = {051604},
  numpages = {4},
  year = {2013},
  month = {May},
  publisher = {American Physical Society},
  doi = {10.1103/PhysRevA.87.051604},
  url = {https://link.aps.org/doi/10.1103/PhysRevA.87.051604}
}

@article{zwerger2003mott,
  title={Mott--Hubbard transition of cold atoms in optical lattices},
  author={Zwerger, Wilhelm},
  journal={Journal of Optics B: Quantum and Semiclassical Optics},
  volume={5},
  number={2},
  pages={S9},
  year={2003},
  publisher={IOP Publishing}
}

@article{jaksch1998cold,
  title={Cold bosonic atoms in optical lattices},
  author={Jaksch, Dieter and Bruder, Christoph and Cirac, Juan Ignacio and Gardiner, Crispin W and Zoller, Peter},
  journal={Physical Review Letters},
  volume={81},
  number={15},
  pages={3108},
  year={1998},
  publisher={APS}
}

@article{hoffmann2009visibility,
  title={Visibility of cold atomic gases in optical lattices for finite temperatures},
  author={Hoffmann, Alexander and Pelster, Axel},
  journal={Physical Review A},
  volume={79},
  number={5},
  pages={053623},
  year={2009},
  publisher={APS}
}

@article{giorgini2008theory,
  title={Theory of ultracold atomic Fermi gases},
  author={Giorgini, Stefano and Pitaevskii, Lev P and Stringari, Sandro},
  journal={Reviews of Modern Physics},
  volume={80},
  number={4},
  pages={1215--1274},
  year={2008},
  publisher={APS}
}

@article{helmes2008mott,
  title={Mott transition of fermionic atoms in a three-dimensional optical trap},
  author={Helmes, RW and Costi, TA and Rosch, A},
  journal={Physical Review Letters},
  volume={100},
  number={5},
  pages={056403},
  year={2008},
  publisher={APS}
}

@article{kumar2016interaction,
  title={Interaction-induced topological and magnetic phases in the Hofstadter-Hubbard model},
  author={Kumar, Pramod and Mertz, Thomas and Hofstetter, Walter},
  journal={Physical Review B},
  volume={94},
  number={11},
  pages={115161},
  year={2016},
  publisher={APS}
}

@article{huang2014extended,
  title={Extended dynamical mean-field study of the Hubbard model with long-range interactions},
  author={Huang, Li and Ayral, Thomas and Biermann, Silke and Werner, Philipp},
  journal={Physical Review B},
  volume={90},
  number={19},
  pages={195114},
  year={2014},
  publisher={APS}
}

@article{baumann2011exploring,
  title={Exploring symmetry breaking at the Dicke quantum phase transition},
  author={Baumann, Kristian and Mottl, Rafael and Brennecke, Ferdinand and Esslinger, Tilman},
  journal={Physical Review Letters},
  volume={107},
  number={14},
  pages={140402},
  year={2011},
  publisher={APS}
}

@article{mottl2012roton,
  title={Roton-type mode softening in a quantum gas with cavity-mediated long-range interactions},
  author={Mottl, Rafael and Brennecke, Ferdinand and Baumann, Kristian and Landig, Renate and Donner, T and Esslinger, Tilman},
  journal={Science},
  volume={336},
  number={6088},
  pages={1570--1573},
  year={2012},
  publisher={American Association for the Advancement of Science}
}

@article{klinder2015dynamical,
  title={Dynamical phase transition in the open Dicke model},
  author={Klinder, Jens and Ke{\ss}ler, Hans and Wolke, Matthias and Mathey, Ludwig and Hemmerich, Andreas},
  journal={Proceedings of the National Academy of Sciences},
  volume={112},
  number={11},
  pages={3290--3295},
  year={2015},
  publisher={National Academy of Sciences}
}

@article{landig2015measuring,
  title={Measuring the dynamic structure factor of a quantum gas undergoing a structural phase transition},
  author={Landig, Renate and Brennecke, Ferdinand and Mottl, Rafael and Donner, Tobias and Esslinger, Tilman},
  journal={Nature Communications},
  volume={6},
  number={1},
  pages={7046},
  year={2015},
  publisher={Nature Publishing Group UK London}
}

@article{camacho2017quantum,
  title={Quantum simulation of competing orders with fermions in quantum optical lattices},
  author={Camacho-Guardian, Arturo and Paredes, Rosario and Caballero-Ben{\'\i}tez, Santiago F},
  journal={Physical Review A},
  volume={96},
  number={5},
  pages={051602},
  year={2017},
  publisher={APS}
}

@article{roux2021cavity,
  title={Cavity-assisted preparation and detection of a unitary Fermi gas},
  author={Roux, Kevin and Helson, Victor and Konishi, Hideki and Brantut, Jean-Philippe},
  journal={New Journal of Physics},
  volume={23},
  number={4},
  pages={043029},
  year={2021},
  publisher={IOP Publishing}
}

@article{konishi2021universal,
  title={Universal pair polaritons in a strongly interacting Fermi gas},
  author={Konishi, Hideki and Roux, Kevin and Helson, Victor and Brantut, Jean-Philippe},
  journal={Nature},
  volume={596},
  number={7873},
  pages={509--513},
  year={2021},
  publisher={Nature Publishing Group UK London}
}

@article{zwettler2025nonequilibrium,
  title={Nonequilibrium Dynamics of Long-Range Interacting Fermions},
  author={Zwettler, Timo and Del Pace, Giulia and Marijanovic, Filip and Chattopadhyay, Sambuddha and B{\"u}hler, Tabea and Halati, C-M and Skolc, Luka and Tolle, Luisa and Helson, Victor and Bolognini, Gaia and others},
  journal={Physical Review X},
  volume={15},
  number={2},
  pages={021089},
  year={2025},
  publisher={APS}
}

@article{carl2023phases,
  title={Phases, instabilities and excitations in a two-component lattice model with photon-mediated interactions},
  author={Carl, Leon and Rosa-Medina, Rodrigo and Huber, Sebastian D and Esslinger, Tilman and Dogra, Nishant and Dubcek, Tena},
  journal={Physical Review Research},
  volume={5},
  number={3},
  pages={L032003},
  year={2023},
  publisher={APS}
}

@article{hoang2002coherent,
  title={Coherent potential approximation for charge ordering in the extended Hubbardmodel},
  author={Hoang, Anh-Tuan and Thalmeier, P},
  journal={Journal of Physics: Condensed Matter},
  volume={14},
  number={26},
  pages={6639},
  year={2002},
  publisher={IOP Publishing}
}

@article{micnas1990superconductivity,
  title={Superconductivity in narrow-band systems with local nonretarded attractive interactions},
  author={Micnas, R and Ranninger, J and Robaszkiewicz, St},
  journal={Reviews of Modern Physics},
  volume={62},
  number={1},
  pages={113},
  year={1990},
  publisher={APS}
}

@article{hirsch1984charge,
  title={Charge-density-wave to spin-density-wave transition in the extended Hubbard model},
  author={Hirsch, JE},
  journal={Physical Review Letters},
  volume={53},
  number={24},
  pages={2327},
  year={1984},
  publisher={APS}
}

@article{lin1986condensation,
  title={Condensation transition in the one-dimensional extended Hubbard model},
  author={Lin, HQ and Hirsch, JE},
  journal={Physical Review B},
  volume={33},
  number={12},
  pages={8155},
  year={1986},
  publisher={APS}
}

@article{aichhorn2004charge,
  title={Charge ordering in extended Hubbard models: Variational cluster approach},
  author={Aichhorn, Markus and Evertz, Hans Gerd and von der Linden, Wolfgang and Potthoff, Michael},
  journal={Physical Review B},
  volume={70},
  number={23},
  pages={235107},
  year={2004},
  publisher={APS}
}

@article{merino2005quantum,
  title={Quantum melting of charge order due to frustration in two-dimensional quarter-filled systems},
  author={Merino, Jaime and Seo, Hitoshi and Ogata, Masao},
  journal={Physical Review B},
  volume={71},
  number={12},
  pages={125111},
  year={2005},
  publisher={APS}
}

@article{fratini2009unconventional,
  title={Unconventional metallic conduction in two-dimensional Hubbard-Wigner lattices},
  author={Fratini, Simone and Merino, J},
  journal={Physical Review B},
  volume={80},
  number={16},
  pages={165110},
  year={2009},
  publisher={APS}
}

@article{davoudi2006nearest,
  title={Nearest-neighbor repulsion and competing charge and spin order in the extended Hubbard model},
  author={Davoudi, B and Tremblay, A-MS},
  journal={Physical Review B},
  volume={74},
  number={3},
  pages={035113},
  year={2006},
  publisher={APS}
}

@article{vojta1999charge,
  title={Charge-order transition in the extended Hubbard model on a two-leg ladder},
  author={Vojta, Matthias and Hetzel, RE and Noack, RM},
  journal={Physical Review B},
  volume={60},
  number={12},
  pages={R8417},
  year={1999},
  publisher={APS}
}

@article{vojta2001phase,
  title={Phase diagram of the quarter-filled extended Hubbard model on a two-leg ladder},
  author={Vojta, Matthias and H{\"u}bsch, Arnd and Noack, RM},
  journal={Physical Review B},
  volume={63},
  number={4},
  pages={045105},
  year={2001},
  publisher={APS}
}

@article{amaricci2010extended,
  title={Extended hubbard model: Charge ordering and wigner-mott transition},
  author={Amaricci, Adriano and Camjayi, Alberto and Haule, Kristjan and Kotliar, G and Tanaskovi{\'c}, D and Dobrosavljevi{\'c}, V},
  journal={Physical Review B},
  volume={82},
  number={15},
  pages={155102},
  year={2010},
  publisher={APS}
}

@article{camjayi2008coulomb,
  title={Coulomb correlations and the Wigner--Mott transition},
  author={Camjayi, A and Haule, K and Dobrosavljevi{\'c}, V and Kotliar, G},
  journal={Nature Physics},
  volume={4},
  number={12},
  pages={932--935},
  year={2008},
  publisher={Nature Publishing Group UK London}
}

@article{tong2004charge,
  title={Charge ordering and phase separation in the infinite dimensional extended Hubbard model},
  author={Tong, Ning-Hua and Shen, Shun-Qing and Bulla, Ralf},
  journal={Physical Review B—Condensed Matter and Materials Physics},
  volume={70},
  number={8},
  pages={085118},
  year={2004},
  publisher={APS}
}

@article{merino2013emergent,
  title={Emergent heavy fermion behavior at the Wigner-Mott transition},
  author={Merino, Jaime and Ralko, Arnaud and Fratini, Simone},
  journal={Physical Review Letters},
  volume={111},
  number={12},
  pages={126403},
  year={2013},
  publisher={APS}
}

@article{ayral2013screening,
  title={Screening and nonlocal correlations in the extended Hubbard model from self-consistent combined GW and dynamical mean field theory},
  author={Ayral, Thomas and Biermann, Silke and Werner, Philipp},
  journal={Physical Review B},
  volume={87},
  number={12},
  pages={125149},
  year={2013},
  publisher={APS}
}

@article{van2014beyond,
  title={Beyond extended dynamical mean-field theory: Dual boson approach to the two-dimensional extended Hubbard model},
  author={Van Loon, Erik GCP and Lichtenstein, Alexander I and Katsnelson, Mikhail I and Parcollet, Olivier and Hafermann, Hartmut},
  journal={Physical Review B},
  volume={90},
  number={23},
  pages={235135},
  year={2014},
  publisher={APS}
}

@article{klinder2015observation,
  title={Observation of a superradiant Mott insulator in the Dicke-Hubbard model},
  author={Klinder, Jens and Ke{\ss}ler, Hans and Bakhtiari, M Reza and Thorwart, M and Hemmerich, Andreas},
  journal={Physical Review Letters},
  volume={115},
  number={23},
  pages={230403},
  year={2015},
  publisher={APS}
}

@article{bloch2008many,
  title={Many-body physics with ultracold gases},
  author={Bloch, Immanuel and Dalibard, Jean and Zwerger, Wilhelm},
  journal={Reviews of Modern Physics},
  volume={80},
  number={3},
  pages={885--964},
  year={2008},
  publisher={APS}
}

@article{tolle2025steady,
  title={Steady state diagram of interacting fermionic atoms coupled to dissipative cavities},
  author={Tolle, Luisa and Sheikhan, Ameneh and Giamarchi, Thierry and Kollath, Corinna and Halati, Catalin-Mihai},
  journal={arXiv preprint arXiv:2509.07469},
  year={2025}
}

@article{tolle2025fluctuation,
  title={Fluctuation-Induced Bistability of Fermionic Atoms Coupled to a Dissipative Cavity},
  author={Tolle, Luisa and Sheikhan, Ameneh and Giamarchi, Thierry and Kollath, Corinna and Halati, Catalin-Mihai},
  journal={Physical Review Letters},
  volume={134},
  number={13},
  pages={133602},
  year={2025},
  publisher={APS}
}

@article{Buhler_2025_Microscopy,
  title={Microscopy of cavity-induced density-wave ordering in ultracold gases},
  author={B{\"u}hler, Tabea and Fabre, Aur{\'e}lien and Bolognini, Gaia and Xue, Zeyang and Zwettler, Timo and Del Pace, Giulia and Brantut, Jean-Philippe},
  journal={arXiv preprint arXiv:2511.08510},
  year={2025}
}

@article{black2003observation,
  title={Observation of Collective Friction Forces due to Spatial Self-Organization of Atoms: From Rayleigh to Bragg Scattering},
  author={Black, Adam T and Chan, Hilton W and Vuleti{\'c}, Vladan},
  journal={Physical Review Letters},
  volume={91},
  number={20},
  pages={203001},
  year={2003},
  publisher={APS}
}

@article{liao2018theoretical,
  title={Theoretical exploration of competing phases of lattice Bose gases in a cavity},
  author={Liao, Renyuan and Chen, Huang-Jie and Zheng, Dong-Chen and Huang, Zhi-Gao},
  journal={Physical Review A},
  volume={97},
  number={1},
  pages={013624},
  year={2018},
  publisher={APS}
}

@article{mermin1966absence,
  title={Absence of ferromagnetism or antiferromagnetism in one-or two-dimensional isotropic Heisenberg models},
  author={Mermin, N David and Wagner, Herbert},
  journal={Physical Review Letters},
  volume={17},
  number={22},
  pages={1133},
  year={1966},
  publisher={APS}
}

@article{hohenberg1967existence,
  title={Existence of long-range order in one and two dimensions},
  author={Hohenberg, Pierre C},
  journal={Physical Review},
  volume={158},
  number={2},
  pages={383},
  year={1967},
  publisher={APS}
}

@article{palle2021physical,
  title={Physical limitations of the Hohenberg--Mermin--Wagner theorem},
  author={Palle, Grgur and Sunko, Denis Karl},
  journal={Journal of Physics A: Mathematical and Theoretical},
  volume={54},
  number={31},
  pages={315001},
  year={2021},
  publisher={IOP Publishing}
}

@misc{nhr,
    howpublished = {www.nhr-verein.de/unsere-partner}
}

@misc{gauss,
    howpublished = {www.gauss-centre.eu}
}
\end{document}